\documentclass[final,3p,twocolumn,times]{elsarticle}

\usepackage{amssymb}
\usepackage{amsmath}

\journal{Physica B}

\begin{document}

\begin{frontmatter}

\title{Anderson Localization in Bi-layer Array with Compositional Disorder: Conventional Photonic Crystals versus Metamaterials}

\author[IFUAP]{F.~M.~Izrailev}
\ead{izrailev@sirio.ifuap.buap.mx}

\address[IFUAP]{Instituto de F\'{\i}sica, Universidad Aut\'{o}noma de Puebla, Apartado Postal J-48, Puebla 72570, M\'{e}xico}

\author[ICUAP]{N.~M.~Makarov\corref{CorAuth}}
\ead{makarov@siu.buap.mx}
\cortext[CorAuth]{Corresponding author}

\address[ICUAP]{Instituto de Ciencias, Universidad Aut\'{o}noma de Puebla, Privada 17 Norte No. 3417, Col. San Miguel Hueyotlipan, Puebla 72050, M\'{e}xico}

\author[IFUAP]{E.~J.~Torres-Herrera}

\begin{abstract}
The localization length has been derived for one-dimensional bi-layered structures with random perturbations in the refractive indices for each type of layers. Main attention is paid to the comparison between conventional materials and those consisting of mixed right-hand and left-hand materials. It is shown that the localization length is described by the universal expression for both cases. The analytical results are confirmed by numerical simulations.
\end{abstract}


\end{frontmatter}


\textbf{Introduction} -- In many fields of fundamental and applied physics much attention is paid to the wave propagation and electron transport in one-dimensional periodic structures with elementary cells consisting of two components (see, e.g. \cite{MS08} and references therein). For example, it could be two optical or electromagnetic materials, or the pair of quantum wells and barriers in electronics. The interest to such bi-layer structures is due to various applications for which one needs to know how to create materials,
metamaterials, or semiconductor superlattices with given transmission properties.

One of the important problems that still remains open, is the role of a disorder that cannot be avoided in experimental devices due to fluctuations of the width of layers or due to variations of the medium parameters, such as the dielectric constant, the magnetic permeability, or the barrier hight for electrons. In spite of remarkable progress in this field, the majority of studies of the wave (electron) propagation through random structures are based on various numerical methods, with an additional assumption of rapidly decaying correlations \cite{MCMM93,So00,SSS01,Po03,VM03,Eo06,DZ06,No07,Po07,Ao07,NML07,NML08}. As for the analytical results, they are mainly obtained either for systems with a white-noise disorder \cite{BW85}, or for the patterns with correlated disorder, however, with delta-like potential wells \cite{IK99} or barriers \cite{IKU01,HIT08}.

In this paper we derive the unique analytical expression for the localization length $L_{loc}$ that is valid for a discrete bi-layer structures with weakly disordered refractive indices of both basic slabs. In contrast with the general approach, see \cite{IM09}, here we are interested in a particular problem of comparison of conventional photonic crystals with metamaterials.

\textbf{Model} -- We consider the propagation of an electromagnetic wave of frequency $\omega$ through an infinite array (stack) of two alternating $a$ and $b$ layers (slabs). Every kind of slabs is respectively specified by the dielectric permittivity $\varepsilon_{a,b}$, magnetic permeability $\mu_{a,b}$,
refractive index $n_{a,b}=\sqrt{\varepsilon_{a,b}\mu_{a,b}}$, impedance $Z_{a,b}=\mu_{a,b}/n_{a,b}$ and wave number $k_{a,b}=\omega n_{a,b}/c$. We address two cases when $a$ slabs contain conventional right-handed (RH) optic material, while $b$ layers are composed of either RH or left-handed (LH) material. The combination of RH-RH slabs is caled \emph{homogeneous} stack whereas the array of RH-LH layers is called \emph{mixed} stack. Following Ref.~\cite{Ao07}, a disorder is incorporated via the dielectric constants only (\emph{compositional disorder}), so that
\begin{subequations}\label{na-nb}
\begin{eqnarray}
&&\mu_a=1,\quad n_a(n)=1+\eta_a(n); \label{na}\\
&&\mu_b=\pm1,\quad n_b(n)=\pm[1+\eta_b(n)].\label{nb}
\end{eqnarray}
\end{subequations}
Here integer $n$ enumerates the elementary $(ab)$ cells. The upper sign is related to RH material while the lower one is associated with LH media. Every alternating slab has the constant width $d_a$ or $d_b$, respectively.

Without disorder, $\eta_{a,b}(n)=0$, all layers are perfectly matched ($Z_a=Z_b$) and the stack is equivalent to the homogeneous medium with the refractive index $\overline{n}$,
\begin{equation}\label{UnpDR}
\kappa=\omega\overline{n}/c,\qquad
\overline{n}=|d_a\pm d_b|/(d_a+d_b),
\end{equation}
with no gaps in the linear spectrum.
Remarkably, in the ideal mixed stack ($\varepsilon_a=\mu_a=1$,
$\varepsilon_b=\mu_b=-1$, $Z_a=Z_b=1$) with equal slab widths, the phase velocity $c/\overline{n}$ diverges.
Therefore, our consideration is meaningful only when two layers have different widths $d_a\neq d_b$.

The random sequences $\eta_{a}(n)$ and $\eta_{b}(n)$ are statistically homogeneous with the zero mean, $\langle\eta_{a,b}(n)\rangle=0$, and binary correlation functions defined by
\begin{subequations}\label{CorrDef}
\begin{eqnarray}
&&\langle\eta_a(n)\eta_a(n')\rangle=
\sigma_{a}^2\,K_a(n-n')\,,\\
&&\langle\eta_b(n)\eta_b(n')\rangle=
\sigma_{b}^2\,K_b(n-n')\,,\\
&&\langle\eta_a(n)\eta_b(n')\rangle=
\sigma_{ab}^2\,K_{ab}(n-n').
\end{eqnarray}
\end{subequations}
The average $\langle ... \rangle$ is performed over the whole array or due to the ensemble averaging, that is assumed to be the same. The auto-correlators $K_{a}(r)$ and $K_{b}(r)$ as well as the inter-correlator $K_{ab}(r)$ are normalized to one: $K_{a}(0)=K_{b}(0)=K_{ab}(0)=1$. The variances
$\sigma^2_{a}$ and $\sigma^2_{b}$ are obviously positive, however, the term $\sigma_{ab}^2$ can be of arbitrary value. We assume the disorder to be \emph{weak}, $(k_{a,b}d_{a,b}\sigma_{a,b})^2\ll1$,
that allows us to develop a proper perturbation theory. In this case all transport properties are entirely determined by the randomness power spectra $\mathcal{K}_a(k)$, $\mathcal{K}_b(k)$, and $\mathcal{K}_{ab}(k)$, defined by the relations
\begin{equation}\label{FT-K}
\mathcal{K}(k)=\sum_{r=-\infty}^{\infty}K(r)\exp(-ikr)=
1+2\sum_{r=1}^{\infty}K(r)\cos(kr).
\end{equation}
By definition \eqref{CorrDef}, all the correlators $K_{a}(r)$, $K_{b}(r)$ and $K_{ab}(r)$ are real and even functions of the difference $r=n-n'$ between cell indices. Because of this fact and due to their positive normalization, the corresponding Fourier transforms $\mathcal{K}_a(k)$, $\mathcal{K}_b(k)$ and $\mathcal{K}_{ab}(k)$ are real, even and non-negative functions of the dimensionless lengthwise wave number $k$.

Within every $a$ or $b$ layer, the electric field of the wave obeys the 1D Helmholtz equation with two boundary conditions at the interfaces between neighboring slabs,
\begin{subequations}\label{WaveEqBC}
\begin{eqnarray}
\left(\frac{d^2}{dx^2}+k_{a,b}^2\right)\psi_{a,b}(x)=0,
\label{WaveEq-ab}\\
\psi_a(x_i)=\psi_b(x_i),\quad
\mu_a^{-1}\psi'_a(x_i)=\mu_b^{-1}\psi'_b(x_i).
\label{BC}
\end{eqnarray}
\end{subequations}
The $x$ axis is directed along the array of bi-layers, and
$x=x_{i}$ stands for the interface coordinate.

\textbf{Hamiltonian map approach}~\cite{IK99,IKU01,HIT08} -- The solution of Eq.~\eqref{WaveEqBC} can be presented as the recurrent relations for the wave function $\psi_{an}=Q_n$ and its derivative $(c/\omega)\psi'_{an}=P_n$ at the two opposite edges of the $n$th elementary $(a,b)$ cell,
\begin{equation}\label{map-QP}
Q_{n+1}=A_nQ_n+B_nP_n,\quad P_{n+1}=-C_nQ_n+D_nP_n.
\end{equation}
The factors $A_n$, $B_n$, $C_n$, $D_n$ read
\begin{subequations}\label{ABCDn}
\begin{eqnarray}
A_n&=&\cos\widetilde{\varphi}_a\cos\widetilde{\varphi}_b-
Z_a^{-1}Z_b\sin\widetilde{\varphi}_a\sin\widetilde{\varphi}_b,\\
B_n&=&Z_a\sin\widetilde{\varphi}_a\cos\widetilde{\varphi}_b+
Z_b\cos\widetilde{\varphi}_a\sin\widetilde{\varphi}_b,\\
C_n&=&Z_a^{-1}\sin\widetilde{\varphi}_a\cos\widetilde{\varphi}_b+
Z_b^{-1}\cos\widetilde{\varphi}_a\sin\widetilde{\varphi}_b,\\
D_n&=&\cos\widetilde{\varphi}_a\cos\widetilde{\varphi}_b-
Z_aZ_b^{-1}\sin\widetilde{\varphi}_a\sin\widetilde{\varphi}_b.
\end{eqnarray}
\end{subequations}
They depend on the cell index $n$ due to the random refractive indices \eqref{na-nb}, which enter the impedances $Z_{a,b}$, as well as due to the phase shifts
\begin{eqnarray}\label{phi-ab}
&&\widetilde{\varphi}_{a,b}(n)=\varphi_{a,b}[1+\eta_{a,b}(n)],
\nonumber\\
&&\varphi_{a}=\omega d_a/c,\qquad\varphi_{b}=\pm\omega d_b/c.
\end{eqnarray}
It is noteworthy to emphasize that the recurrent relations \eqref{map-QP} can be treated as the Hamiltonian map of trajectories in the phase space $(Q,P)$ with discrete time $n$ for a linear oscillator subjected to time-depended parametric force.

Without disorder $\eta_{a,b}(n)=0$, the factors \eqref{ABCDn} do not depend on the time $n$. Therefore, the trajectory $Q_n,P_n$ creates a circle in the phase space $(Q,P)$ that is an image of the unperturbed motion,
\begin{eqnarray}\label{map-QPunp}
Q_{n+1}=Q_n\cos\gamma+P_n\sin\gamma,\nonumber\\
P_{n+1}=-Q_n\sin\gamma+P_n\cos\gamma.
\end{eqnarray}
The unperturbed phase shift $\gamma$ over a single elementary
$(ab)$ cell is defined as
\begin{equation}\label{gamma-def}
\gamma=\varphi_a+\varphi_b=\omega(d_a\pm d_b)/c,
\end{equation}
that gives $\gamma=\kappa(d_a+d_b)$ due to Eq.~\eqref{UnpDR}. Having the circle, it is suitable to pass to \emph{action-angle variables} $R_n$ and $\theta_n$ via the standard transformation
\begin{equation}\label{QP-RTheta}
Q_n=R_n\cos\theta_n,\qquad P_n=R_n\sin\theta_n.
\end{equation}
By direct substitution of Eq.~\eqref{QP-RTheta} into the map \eqref{map-QPunp}, one can reveal that for the unperturbed trajectory the radius $R_n$ is conserved, while its phase $\theta_n$ changes by the \emph{Bloch phase} $\gamma$ in one step of time $n$,
\begin{equation}\label{map-RThetaUnp}
R_{n+1}=R_n,\qquad \theta_{n+1}=\theta_n-\gamma.
\end{equation}

Evidently, a weak random perturbation results in a small distortion of the circle \eqref{map-RThetaUnp} that can be evaluated in the following way. First, in the initial map \eqref{map-QP} we expand the factors \eqref{ABCDn} up to the second order in the perturbation parameters $\eta_{a,b}(n)\ll1$ entering the impedances $Z_{a,b}(n)$ and phase shifts $\widetilde{\varphi}_{a,b}(n)$. After getting the perturbed map for $Q_n$ and $P_n$, we pass to action-angle variables $R_n$ and $\theta_n$ with the use of Eq.~\eqref{QP-RTheta}. All these quite cumbersome calculations allow us to derive the perturbed map for the radius $R_n$ and angle $\theta_n$ keeping linear and quadratic terms in the perturbation:
\begin{subequations}\label{map-WD}
\begin{eqnarray}
R_{n+1}^2/R_n^2=1+\eta_{a}(n)V_{a}(\theta_n)+ \eta_{b}(n)V_{b}(\theta_n)\nonumber\\
+\eta_{a}^2(n)W_{a}+\eta_{b}^2(n)W_{b}+
\eta_{a}(n)\eta_{b}(n)W_{ab},\label{mapR-WD}\\
\theta_{n+1}-\theta_n+\gamma=\eta_{a}(n)U_{a}(\theta_n)+ \eta_{b}(n)U_{b}(\theta_n)\label{mapTheta-WD}.
\end{eqnarray}
\end{subequations}
Here the functions standing with random variables $\eta_{a,b}(n)$ are described by the expressions:
\begin{subequations}\label{VW}
\begin{eqnarray}
&&V_{a}(\theta_n)=-2\sin\varphi_{a}\sin(2\theta_n-\varphi_{a}),\\
&&V_{b}(\theta_n)=-2\sin\varphi_{b}\sin(2\theta_n-\gamma-\varphi_{a}),\\
&&W_{a}=2\sin^2\varphi_{a},\quad W_{b}=2\sin^2\varphi_{b},\\
&&W_{ab}=4\sin\varphi_{a}\sin\varphi_{b}\cos\gamma;\\
&&U_{a}(\theta_n)=-\sin\varphi_{a}\cos(2\theta_n-\varphi_{a}),\\
&&U_{b}(\theta_n)=-\sin\varphi_{b}\cos(2\theta_n-\gamma-\varphi_{a}).
\end{eqnarray}
\end{subequations}
Note that in Eqs.~\eqref{VW} we keep only the terms that contribute to the localization length $L_{loc}$ in the first non-vanishing order of approximation. Since in Eq.~\eqref{mapR-WD} the factors $V_{a,b}$ containing $\theta_n$ are always multiplied by $\eta_{a,b}(n)$, only linear terms in these perturbation parameters are needed in the complementing recurrent relation \eqref{mapTheta-WD} for the angle $\theta_n$. Relations \eqref{map-WD} constitute the complete set of equations in order to derive the localization length of the system under consideration.

\textbf{Localization length} -- We define the localization length $L_{loc}$ via the Lyapunov exponent $\lambda$ \cite{IKT95,IRT98},
\begin{equation}\label{LyapDef}
\frac{d_a+d_b}{L_{loc}}\equiv\lambda=\frac{1}{2}
\Big\langle\overline{\ln\left(\frac{R_{n+1}}{R_n}\right)^2}\Big\rangle.
\end{equation}
The average $\langle ab\rangle$ is performed over the disorder parameters
$\eta_{a,b}(n)$, and the average $\overline{ab}$ is carried out over the random phase $\theta_n$. Now we substitute the recurrent relation \eqref{mapR-WD} into the definition \eqref{LyapDef} and expand the logarithm within the quadratic approximation in the perturbation parameters. Then we perform the averaging over both the disorder and rapid phase assuming the distribution of $\theta_n$ to be homogenous within the first order of approximation. After some algebra we arrive at the final expression for the Lyapunov exponent:
\begin{equation*}
\lambda=
\frac{1}{2}\sigma_{a}^2
\mathcal{K}_a(2\gamma)\sin^2\varphi_{a}
+\frac{1}{2}\sigma_{b}^2
\mathcal{K}_b(2\gamma)\sin^2\varphi_{b}
\end{equation*}
\begin{equation}\label{LyapFin}
+\sigma_{ab}^2\mathcal{K}_{ab}(2\gamma)
\sin\varphi_{a}\sin\varphi_{b}\cos\gamma.
\end{equation}
Note that the Eq.~\eqref{LyapFin} is expectedly symmetric with respect to the permutation of slab indices $a\leftrightarrow b$.

In accordance with Eq.~\eqref{LyapFin} the Lyapunov exponent $\lambda$ (the inverse localization length $L^{-1}_{loc}$) consists of three terms. The first two terms are contributed respectively by the correlations between solely $a$ or solely $b$ slabs. Therefore, these terms contain the \emph{auto-correlators} $\mathcal{K}_a(2\gamma)$ or $\mathcal{K}_b(2\gamma)$. The third term emerges due to the inter-correlations between two, $a$ and $b$, disorders. It includes the \emph{inter-correlator} $\mathcal{K}_{ab}(2\gamma)$.

Eq.~\eqref{LyapFin} manifests that the only difference for the homogeneous RH-RH and mixed RH-LH stacks is due to the sign
in the phase shift $\varphi_{b}$. This affects the value \eqref{gamma-def} of the Bloch phase $\gamma$ and the sign at the third inter-correlation term. One can also see that the Lyapunov exponent typically obeys the \emph{conventional} frequency dependence
\begin{equation}\label{LyapOmega2}
\lambda\propto L_{loc}^{-1}\propto\omega^2\qquad \mathrm{when}\quad\omega\to0.
\end{equation}
However, specific correlations in the potential, taken into account in Eq.~\eqref{LyapFin}, may result in a quite unusual $\omega-$dependence, see \cite{IM09}.

The Lyapunov exponent $\lambda(\omega)$ exhibits the \emph{Fabry-Perot resonances} associated with multiple reflections inside $a$ or $b$ slabs from the interfaces. As is known, they appear when the width $d_{a,b}$ of corresponding $a$ or $b$ layer equals to an integer multiple of half of the wavelength $2\pi c/\omega$ inside the layer,
\begin{equation}\label{BiL-FPab}
\omega/c=s_a\pi/d_a\quad\mathrm{or}\quad
\omega/c=s_b\pi/d_b,\quad
s_{a,b}=1,2,3,\dots.
\end{equation}
At the resonances the factor $\sin\varphi_a$ or $\sin\varphi_b$ in Eq.~\eqref{LyapFin} vanishes, resulting in the resonance increase of the localization length $L_{loc}$ and consequently, in \emph{suppression of the localization}. In the special case when the ratio of slab widths $d_a$ and $d_b$ turns out to be a rational number, $d_a/d_b=s_a/s_b$, some resonances from different types of layers coincide and give rise to the divergence of the localization length $L_{loc}(\omega)$. Remarkably, the Fabry-Perot resonance is quite \emph{broad} because it is caused by vanishing of smooth trigonometric functions \cite{LIMKS09}.

Of special interest are long-range correlations leading to the
divergence or significant decrease of the localization length $L_{loc}(\omega)$ in the controlled frequency window. This effect is similar to that found in more simple 1D models with correlated disorder. In our model this effect is due to a possibility to have the vanishing values of all Fourier transforms, $\mathcal{K}_{a}=\mathcal{K}_{b}=\mathcal{K}_{ab}=0$, in some intervals of frequency $\omega$. For example, one can artificially construct an array of random bi-layers with such power spectra that abruptly vanish within a prescribed interval of $\omega$, resulting in the \emph{divergence} of the localization length \cite{IK99,KIKS00,IKU01,HIT08}. On the contrary, with the use of specific correlations one can decrease the localization length, and significantly \emph{enhance} the localization \cite{KIK08}.

Finally, let us consider the particular case of the white-noise disorders for $a$ and $b$ slabs,
\begin{equation}\label{Corr-FR}
\sigma_{ab}^2=0,\qquad\mathcal{K}_a(k)=\mathcal{K}_b(k)=1.
\end{equation}
Here, the Lyapunov exponent and the inverse localization length turn out to be  exactly the \emph{same} for both homogeneous RH-RH and mixed RH-LH stack-structures,
\begin{equation}\label{LyapWN}
\lambda=\frac{d_a+d_b}{L_{loc}}=\frac{1}{2}
(\sigma_a^2\sin^2\varphi_{a}+\sigma_b^2\sin^2\varphi_{b}).
\end{equation}
The numerical results shown in Fig.~1 are obtained with the use of Eqs.~\eqref{map-QP}, without any approximation.
\begin{figure}[h]
\begin{center}
\includegraphics[scale=0.6]{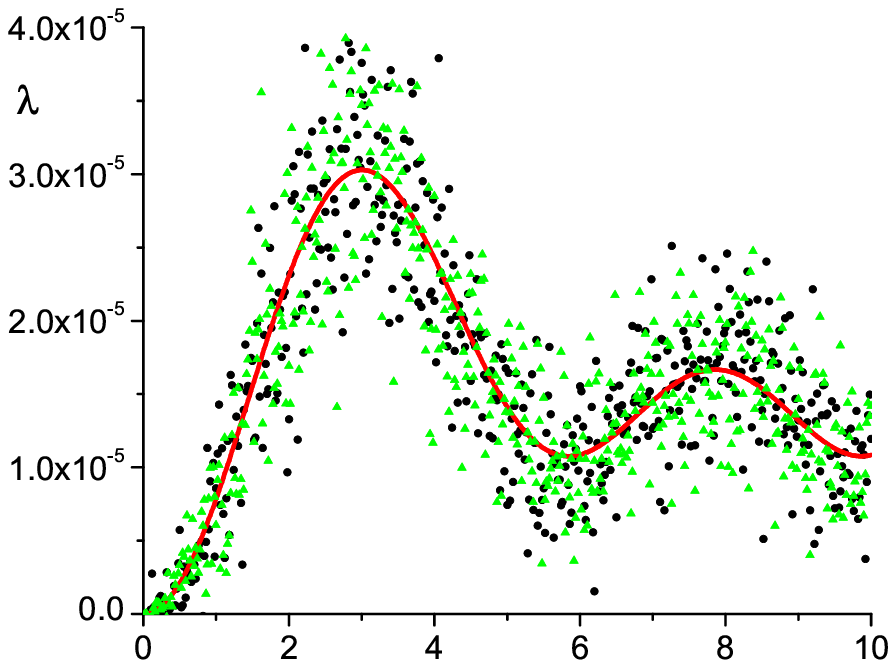}
\end{center}
\end{figure}
\begin{figure}[h]
\vspace{-1.5cm}
\begin{center}
\includegraphics[scale=0.6]{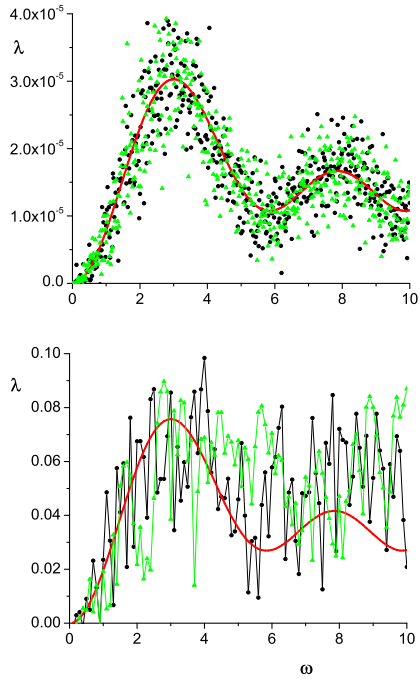}
\caption{Lyapunov exponent versus frequency. Top: RH-RH (circles) and RH-LH (triangles) media for $\sigma_a\approx\sigma_b\approx 0.006\,, d_a=0.6\,, d_b=0.4\,, c=1$ and the length of sequence is $N=10^6$. Bottom: the same for $\sigma_a\approx\sigma_b\approx 0.3$ and the $N=100$. Smooth curve depicts Eq.~\eqref{LyapWN}.}
\end{center}
\end{figure}
In the higher panel one can see that for a very long sample and weak disorder the analytical expression \eqref{LyapWN} perfectly corresponds to the data, apart from fluctuations. For each case only one realization of the disorder was used. The fluctuations can be smoothed out by an additional ensemble averaging. In order to see whether our predictions can be used in experiment, we also show in the lower panel the data for a very short sample and very strong disorder. As one can see, the analytical result is also valid for small frequencies, and gives the qualitatively correct Lyapunov exponent for large values of $\omega$.

The more detailed comparison for $\omega(d_a+d_b)/c\ll1$ and $\omega(d_a+d_b)/c\gg1$ also shows a nice correspondence. In this respect, a very unusual result obtained in Ref.~\cite{Ao07}, namely, $\lambda\sim\omega^6$, seems to be entirely related to a specific case of equal widths, $d_a=d_b$.

F.M.I acknowledges the support by CONACyT grant No. 80715.



\end{document}